\documentclass[aps,epsfig,twocolumn,showpacs]{revtex4-1}
\usepackage{amssymb}
\usepackage{graphics}
\usepackage{graphicx}
\usepackage{epsfig}
\begin{document}
\title{Pulling short DNA molecules having defects on different locations}
\author{Amar Singh, Navin Singh}
\affiliation{Department of Physics, Birla Institute of Technology \& Science, Pilani - 333 031, 
Rajasthan, India}
\begin{abstract}
We present a study on the role of defects on the stability 
of short DNA molecules. We consider short DNA molecules (16 base pairs) and
investigate the thermal as well as mechanical denaturation of these molecules
in the presence of defects that occurs anywhere in the molecule. 
For the investigation, we consider four different kinds of chains.
Not only the ratio of AT to GC different in these molecules 
but also the distributions of AT and GC along the molecule are different.
With suitable modifications in the statistical model to show the defect
in a pair, we investigate the denaturation of short DNA molecules in thermal 
as well as constant force ensemble. In the force ensemble, we pulled the
DNA molecule from each end (keeping other end free) and observed
some interesting features of opening of the molecule in the presence
of defects in the molecule. We calculate the probability of opening of
the DNA molecule in the constant force ensemble to explain the
opening of base pairs and hence the denaturation of molecules 
in the presence of defects.

\end{abstract}
\maketitle

\section{Introduction}

Defects in the DNA molecule play a crucial role in biological processes such as  
replication. This is a known fact that DNA is a long polymeric chain that contains four
different kinds of nitrogenous bases. The allowed pairing in the two complementary
strands follow a simple rule, that is, Adenine (A) can form a hydrogen bond with
Thymine (T) while Guanine (G) can form a hydrogen bond with Cytosine (C)
\cite{watson,stryer}. The hydrogen bonding strength for these two base pairs is not same 
as the AT base pair has two hydrogen bonds while GC base pair has three hydrogen bonds.
The approximate ratio of GC and AT bond strengths varies from 1.2 to 1.5 as mentioned 
by various research groups \cite{pb,campa,chen,saul,voul,erp,weber,pbd2009,alex,zoli}. 
In the absence of the complementary base on the opposite strand, the pairing between 
the two bases is absent. This site is called a \textit{defect site} because of 
an absence of a stable (or non-existing) bonding between these two bases 
on the opposite strands \cite{ns01,ns03}. The presence of defects in DNA is related to 
interesting physics and biochemistry of the molecule. The dynamics of these defects 
may delay the replication process and hence lead to the breathing dynamics of opening of 
the chain \cite{gauth}. It is predicted that in embryonic cells, these delays may 
cause the \textit{cell death} while in mature cells like somatic cells, 
this damage (defect) may  be an initiation step in the development of cancer
\cite{paivi,hensey,vilen,branzei,kaufmann}. These defects are present 
in the DNA based actuators. The role of the defects in the designing 
of molecular motor has been discussed by McCullagh {\it et al} 
\cite{mccullagh}. How the defects affect the melting, elastic and other 
properties are problem of scientific interest.
There are many paths to explore the role of the defects and the damage
repair mechanism in the living cells. Theoretical approach to investigate 
the problem routes via molecular dynamics or model based calculations 
\cite{kafri,amb,joyuex,dudu,kumar,frank,ffalo,macedo,nik}.
Our approach is a model based calculation. We use Peyrard Bishop Dauxois (PBD) 
model \cite{pb} to investigate the thermal and mechanical denaturation of DNA molecules
in presence of defects. The main objective of the current study is to investigate
the effect of density and location of defects on the denaturation of 
DNA molecule.

\begin{figure}[t]
\begin{center}
\includegraphics[height=2.25in,width=2.5in]{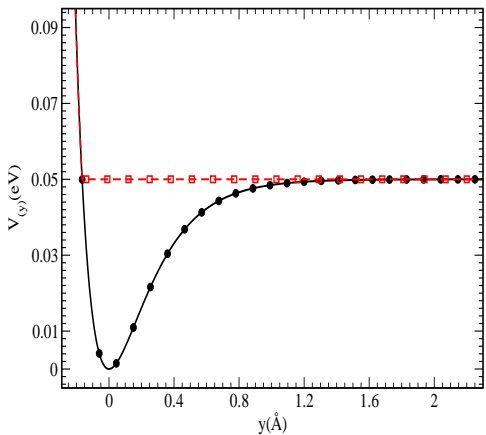}
\caption{\label{fig01}(Color online)
The on-site potential for the defect in a pair is shown by square 
symbol and dashed (red) line. While for the bases in a pair is 
represented by the depth of the potential (solid line with black circle), 
there is no minimum of potential for the defect pair \cite{ns01}.}
\end{center}
\end{figure}

In experiments, researchers synthesis and/or characterize the samples 
to decipher the information stored by that sample. Accordingly, we choose
four samples of DNA molecules each containing 16 base pairs. 
All these samples have different numbers 
and distribution of \textit{AT-GC} pairs. We identify all these molecules 
according to the distribution of base pairs and named them as follows: 
{\it Chain 1: 3'-AAAAAAAAAAAAAAAA-5' } (homogeneous), 
{\it Chain 2: 3'-AGAGAGAGAGAGAGAG-5' } (alternating $AT-GC$ pairs),
{\it Chain 3: 3'-AAAAAAAAGGGGGGGG-5' }(50\%$AT$+50\%$GC$), 
{\it Chain 4: 3'-TCCCTAGACTTAGGGA-5' }(random sequence).
The prime motivation behind the selection of different kinds of 
sequences is to predict the role of defect(s) in the melting or unzipping
of different kinds of DNA molecules. The next task is to introduce the defect in the model.
We have continued from our previous approach \cite{ns01} where the defect 
in the model was introduced via Morse potential that represent the hydrogen bonding. 
If a pair has a defect that means there is an absence of hydrogen bond and 
this feature is reflected from the absence of potential depth  while retaining 
the repulsive part of the potential in order to avoid the crossing of two 
bases in a pair (see Fig. \ref{fig01}). 

We present the work in different sections. In Section \ref{model},
we provide a brief description of the model and methodology used in this work.
The effect of defect(s) on thermal denaturation of dsDNA molecule 
is discussed in Section \ref{melt} while in Section \ref{force}, 
role of the defect(s) on the mechanical unzipping is discussed.
We finally conclude our results in Section \ref{result}.

\section{The model}
\label{model}

For the current investigation we use a statistical model that was proposed by 
Peyrard and Bishop \cite{pb}. The model considers the stretching between the
corresponding bases only. Although the model ignores the helicoidal structure 
of the dsDNA molecule, it has enough details to analyze mechanical behavior 
at the few angstrom scale relevant to molecular-biological events \cite{pbd2009}. 
The Hamiltonian for the considered system of $N$ base pairs unit is 
written as,
\begin{equation}
\label{eqn1}
H = \sum_{i=1}^N\left[\frac{p_i^2}{2m}+ V_M(y_i) \right] +
\sum_{i=1}^{N-1}\left[W_S(y_i,y_{i+1})\right]
\end{equation}
where $y_i$ represents the stretching from the equilibrium position of the hydrogen bonds. 
the first term in the Hamiltonian represents the momentum ($p_i = m${\it\.{y}}).
The $m$ represents the reduced mass of a base pair which is taken to be same for both 
$AT$ and  $GC$ base pairs. The stacking interaction between two consecutive base pairs along the 
chain is represented by, 
\begin{equation}
\label{eqn2}
W_S(y_i,y_{i+1}) = \frac{k}{2}(y_i - y_{i+1})^2[1 + \rho e^{-b(y_i + y_{i+1})}],
\end{equation}
where $k$ represents the single strand elasticity. The anharmonicity in the strand 
elasticity is represented by $\rho$ while $b$ represents its range. The stacking interaction
$W_S(y_i,y_{i+1})$ is independent of the nature of the bases at site $i$ and $i+1$ as 
these parameters are assumed to be independent of sequence heterogeneity. The sequence 
heterogeneity has effect on the stacking interaction along the strand. This can be taken 
care of through the single strand elasticity parameter $k$. One can take the 
variable $k$ according to the distribution of bases along the strand
\cite{as_phy02}. A defect in a pair will modify the electronic distribution around the 
bases hence the stacking parameters. 
However, for the current investigation we settled on the average of this
parameter.

The hydrogen bonding between the two bases in the $i^{th}$ pair is represented by 
the Morse potential. 
\begin{equation}
\label{eqn3}
V_M(y_i) = D_i(e^{-a_iy_i} - 1)^2 
\end{equation}
where $D_i$ represents the potential depth which basically represents the 
bond strength of that pair. The parameter, $a_i$, represents the inverse of 
the width of the potential well. The heterogeneity in the sequence is taken 
care of by the values of $D_i$ and $a_i$. These model parameters should be tuned 
in order to get physical picture of DNA molecule. For the current investigations, 
we choose: $D_{\rm AT} = 0.1 \; {\rm eV}, \; a_{\rm AT} = 4.2 \; {\rm \AA^{-1}}, \;
D_{\rm GC} = 0.15 \; {\rm eV}, \; a_{\rm GC} = 6.3 \; {\rm \AA^{-1}},\;\; \rho = 5.0,
b = 0.35 \; {\rm \AA^{-1}},\; {\rm and} k = 0.021 \; {\rm eV/\AA^{-2}}$.

\begin{figure}[t]
\begin{center}
\includegraphics[height=2in, width=2in]{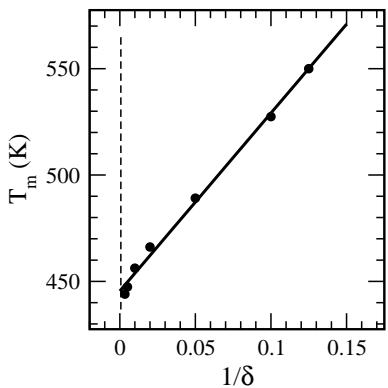}
\caption{\label{fig02} 
The melting temperature $T_m$ calculated for different values of upper 
cut-off, ($\delta$) for homogeneous chain.
The best straight line fit for this plot is found for $1/\delta$. The 
different cut-offs are 8, 10, 20, 50, 100, 200 and 300 \AA.
The model parameters are: $D = 0.1 \; {\rm eV}, \; a = 4.2 \; {\rm \AA^{-1}}, 
\; \rho = 5.0, \; b = 0.35 \; {\rm \AA^{-1}},\; {\rm and} k = 0.021 \; {\rm eV/\AA^{-2}}$.
}
\end{center}
\end{figure}

Thermodynamics of the transition can be investigated by evaluating the expression for 
the partition function. For a sequence of $N$ base pairs with periodic boundary conditions, 
the partition function can be written as:
\begin{equation}
\label{eqn4}
Z = \int_{-\infty}^{\infty} \prod_{i=1}^{N}\left\{dy_idp_i\exp[-\beta H]\right\} = 
Z_pZ_c,
\end{equation}
where $Z_p$ corresponds to the momentum part of the partition function while the $Z_c$
contributes as the configurational part of the partition function. Since 
the momentum part is decoupled in the integration, it can be integrated out as a 
simple Gaussian integral. This will contribute a factor of $(2\pi mk_BT)^{N/2}$ in the 
partition function, where $N$ is the number of base pairs in the chain. 
The calculations of the configurational partition function, $Z_c$, is not straight forward.
This is defined as,
\begin{equation}
\label{eqn5}
Z_c = \int_{-\infty}^{\infty} [\prod_{i=1}^{N-1} dy_i  K(y_i,y_{i+1})]dy_NK(y_N)
\end{equation}
where $K(y_i,y_{i+1}) = \exp\left[-\beta H(y_i,y_{i+1})\right].$
For the homogeneous chain, one can evaluate the partition function by transfer integral (TI)
method by applying the periodic boundary condition \cite{chen}. 
In case of a heterogeneous chain, with open boundary, the configurational part of the 
partition function can be integrated numerically with 
the help of matrix multiplication method \cite{chen,ns03,erp}. The important part of this
integration is the selection of proper cut-offs for the integral appearing in Eq.5
to avoid the divergence of the partition function.
The method to identify the proper cut-off has been discussed by several 
groups \cite{chen,pbd95,erp}.
The calculations done by T.S. van Erp {\it et al} 
show that the upper cut-off will be $\approx$ 144 \AA\ with the 
our model parameters at $T = 600$ K while the lower cut-off is -0.4 \AA.
In the earlier work by Dauxois and Peyrard it was shown that the $T_m$ converges 
rapidly with the upper limit of integration \cite{pbd95}. In that work they
considered an infinite homogeneous chain and solved the partition function using
TI method. For short chains, we 
calculate $T_m$ for different values of upper cut-offs which are shown in 
Fig. \ref{fig02}. From the plot it is clear that the choice of 200 \AA\ is sufficient 
to avoid the divergence of partition function. Thus the configurational space for our
calculations extends from -5 \AA\ to 200 \AA.
Once the  limit of integration has been chosen, the 
task is reduced to discretizing the space to evaluate the integral numerically. 
The space is discretized using the Gaussian quadrature formula. 
In our previous studies \cite{ns03}, we observed that in order to get 
precise value of melting temperature ($T_m$) one has to choose the large 
grid points. We found that 900 is quite sufficient number for this 
purpose. As all matrices in Eq.\ref{eqn6} are identical in nature 
the multiplication is done very efficiently. The thermodynamic 
quantities of interest can be calculated by evaluating the Helmholtz 
free energy of the system. The free energy per base pair is,
\begin{equation}
\label{eqn6}
f(T) = -\frac{1}{2\beta}\ln\left(\frac{2\pi m}{\beta}\right) - 
\frac{1}{N\beta}\ln Z_c; \qquad\qquad \beta = \frac{1}{k_BT}.
\end{equation}
The thermodynamic quantities like specific heat ($C_v$) as a function of temperature
or the applied force can be evaluated by taking the second derivative of the free energy.
The peak in the specific heat corresponds to the melting temperature or the critical
force of the system.

Other quantities such as  the average fraction $\theta(= 1 - \phi)$ of bonded 
(or open) base pairs can be calculated by introducing the dsDNA ensemble(dsDNAE)
\cite{erp} or using the phenomenological approach \cite{campa,ns01}. 
In general, the $\theta$ is defined as,
\begin{equation}
\label{eq7}
\theta = \theta_{\rm ext}\theta_{\rm int} 
\end{equation}
$\theta_{\rm ext}$ is the average fraction of strands forming duplexes, while $\theta_{\rm int}$ 
is the average fraction of unbroken bonds in the duplexes.
The opening of long and short chains are completely different. For long chains, when the 
fraction of open base pairs, $\phi(=1-\theta)$, goes practically from 0 to 1 at 
the melting transition, the two strands are not yet completely separated.
At this point, the majority of the bonds are disrupted and the dsDNA is denaturated, 
but the few bonds still remaining intact, preventing the two strands parting from each other. 
Only at high temperatures will there be a real separation. Therefore for very long chains 
the double strand is always a single macromolecule through the transition, thus one can 
calculate the fraction of intact or broken base pairs only. For short chains, 
the process of single bond disruption and strand dissociation tend to happen 
in the same temperature range. Thus, the computation of both $\theta_{\rm int}$ and 
$\theta_{\rm ext}$ is essential \cite{campa}. The problem of computation of 
$\theta_{\rm ext}$ can be handled efficiently by working in 
dsDNA ensemble (dsDNAE) \cite{erp}.

\section{Thermal melting of the DNA molecule}
\label{melt}

We consider the defects (1-4 in number) and their effect on the melting temperature
of the DNA molecule. Since the nature of each chain is different, the number 
and location of these defects may modify the melting profile of the chain
in different manner.
\vspace{1cm}
\begin{figure}[h]
\begin{center}
\includegraphics[height=2.5in, width=3.2in]{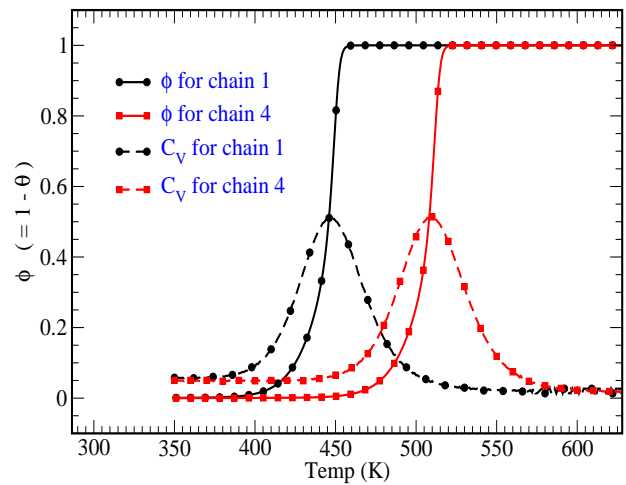}
\caption{\label{fig03}(Color online)
The melting temperature, $T_m$, calculated by specific heat and fraction
of open pair $\phi$ for the {\it chain 1} (homogeneous) and {\it chain 4}
(random). The parameters $p$ and $q$ are adjusted in order to get precise
match with peak in specific heat. The values are $p = 12.0$ and $q = 10.0$.
The value of $C_v$ is scaled to show that the peak position and 50\% 
of the open pairs meet at the same point (temperature).}
\end{center}
\end{figure}
The melting temperature, $T_m$ is calculated by the peak
in the specific heat as well as from $\theta$ as given
in \cite{campa,ns01}. For pure chain, we show the melting profile of the chain
in Fig. \ref{fig03}. 
The melting temperatures for \textit{ chain 1, 2, 3, \& 4} without any defect are 447.5,
508.8, 511.0 and 509.8 K, respectively.
\begin{figure*}[t]
\begin{center}
\includegraphics[height=5.0in, width=6.5in]{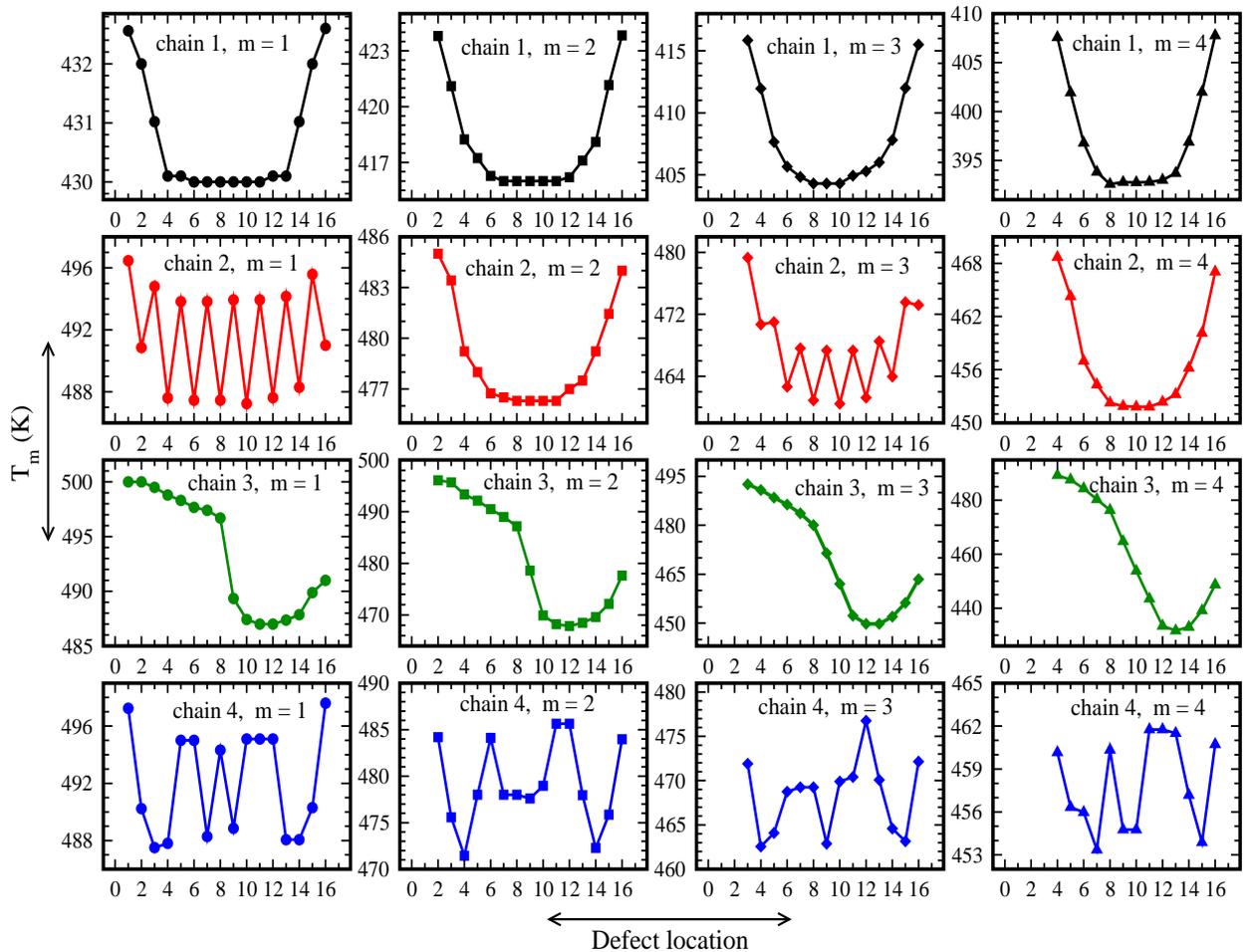}
\caption{\label{fig04} (Color online)
The melting temperature, $T_m$, for all the four chains with different 
numbers as well as the locations of these defects along the DNA molecule. 
Figures in a row are for a chain with different numbers of defects. 
Figures in a column are for different chains with same number of defect(s)
displayed by $m$.}
\end{center}
\end{figure*}
Let us now consider the {\it chain 1} with one defect. When the first site ($3'$- end)
is a defect pair, the melting temperature is about 433 K. The melting temperature
further reduces to 432 K if the $2^{nd}$ pair is a defect pair. However there is
something interesting to note after this. When the location of defect is $4^{th}$ 
pair onward (towards $5'$- end) the melting temperature reduces to 430 K and 
remains constant till we reach $13^{th}$ site. As we reach on the $5'$ end, 
the $T_m$ again increases. This complete cycle displays a necklace kind of 
plot as shown in Fig. \ref{fig04}.
The variation in the melting temperature when the defect is anywhere
between 4 to 13 is negligible. When the number of defects
in the chain is increased from 1 to 4 the width or plateau (where there
is no change in the melting temperature of the molecule) decreases ($\sim$8-12). 

In order to explore more about the nature of denaturation, we investigate 
other chains that have different distribution of base pairs.
Consider \textit{chain 2} with single defect. This chain is having 
alternate $AT/GC$ pairs. As shown in Fig. \ref{fig04}, the symmetry about 
the middle is lost. For this chain, the location of defect site, whether 
it is $AT$ or $GC$ pair, is important. The energy landscape of this chain is not smooth 
over the complete length because of the difference in the dissociation energies
of $AT$ and $GC$ pairs. For single defect that move from position 1 to 16, $T_m$ shows 
a zig-zag pattern and $T_m$ varies between a range of 496 K to 487 K. 
When we consider two consecutive defects in the chain this pattern is lost
because of the loss in the sequence heterogeneity. This can be thought of as 
reviving the homogeneous structure of the DNA molecule with the dual pair
having an average of $AT$ and $GC$ pair's dissociation energies. 
However, the necklace pattern obtained for this case is not as symmetric 
as observed for the {\it chain 1} since the end pairs are not same. 
Remember at $3'$- end there is an $AT$ pair while on $5'$- end, 
there is a $GC$ pair. For this chain, $T_m$ varies between
485 K to 476 K. The zig-zag pattern is retained when three consecutive defects 
are introduced. However, the $T_m$ is lower as compared to the chain 
with one defect. Again with four consecutive defects an asymmetric necklace
is observed with short plateau. 

Let us consider \textit{chain 3}, 
that is having $50AT+50GC$ pairs in the sequence, with one defect. 
In this case, we obtain a hook kind of structure in the plot. 
A sudden drop in the $T_m$ is observed in the middle of the chain 
(on $8\;\&\; 9$ pair) at the interface of \textit{GC} \& \textit{AT} pair.
The smoothness at the interface increases with increase in the number of defects 
in the chain. As the number of defect increases, on the interface the effect of 
presence of $AT$  and $GC$ pairs diminishes. Next is, \textit{chain 4} that is having a 
random distribution of \textit{AT/GC} pairs. Due to random distribution of $AT$ and
$GC$ pairs, the energy landscape is also random. Hence the fluctuation in the 
values of $T_m$ should also be random. This is observed in the figure. 
The random pattern of the plot varies with number of defects in the molecule.
Let us observe the single defect on 4,5,\& 6 sites. While $T_m$ is 488 K for $4^{th}$
site, it is 495 K for $5^{th}$ \& $6^{th}$ sites. For two consecutive 
defects in the molecule, this is averaged to 478 K, \textit{i.e.}. This is
because of the indistinguishability of \textit{AT} and \textit{GC} pair.
Similarly, for three (consecutive) defects in the molecule, the high barrier 
on \textit{11} \& {\it 12} sites, is lost. The pattern observed for this 
chain is closer to the real sequences. 

\section{Force Induced transitions}
\label{force}

The replication process is initiated by the force exerted 
by DNA polymerase on a segment of DNA chain (Owcarzy fragment). The replication
starts at some site which is called replication origin \cite{branzei,kaufmann} 
and the replication fork propagates bidirectionally. The defect or mismatch
pair(s) may slow or stall the replication process. In the case if the 
mismatch repair system does not work properly cell may die \cite{paivi}.
Mathematically one can model the replication as the force applied on
an end of the DNA chain \cite{somen}. Physics of opening of 
chain due to thermal fluctuation and mechanical forces is completely different
\cite{hatch,huguet,ritort}. Thus, the study on the mismatch in the sequence 
and its role in the replication process is an interesting way to 
look into the physics of a complex mechanism. In this section, we 
discuss the force induced unzipping in DNA molecules in presence of defect(s). 
The modified Hamiltonian for the DNA that is pulled mechanically from an
end is,
\begin{equation}
\label{eq8}
H = \sum_{i=1}^N\left[\frac{p_i^2}{2m}+ V_M(y_i) \right] +
\sum_{i=1}^{N-1}\left[W_S(y_i,y_{i+1})\right] -F\cdot y_1
\end{equation}
where the force $F$ is applied on the $1^{st}$ pair \cite{ns03}.
Whereas in thermal denaturation, the opening is due to increase in the 
entropy of the system, for mechanically stretched DNA chain the opening is 
enthalpic. The thermodynamic quantities, of interest, from the modified 
Hamiltonian can be calculated using Eq. \ref{eqn5} \& \ref{eqn6}.
Here we consider the same four chains that we considered for thermal 
denaturation studies. All the base pairs of dsDNA that is kept in 
a thermal bath share equal amount of energy. In the case when the chain 
is pulled from an end, the amount of force decreases from the pulling point 
to the other end of the chain. Thus the location of defect(s) should have 
different impact on the opening of the chain that is subjected to a 
mechanical pull from an end. 
\begin{figure*}[t]
\begin{center}
\includegraphics[height=5in, width=6.5in]{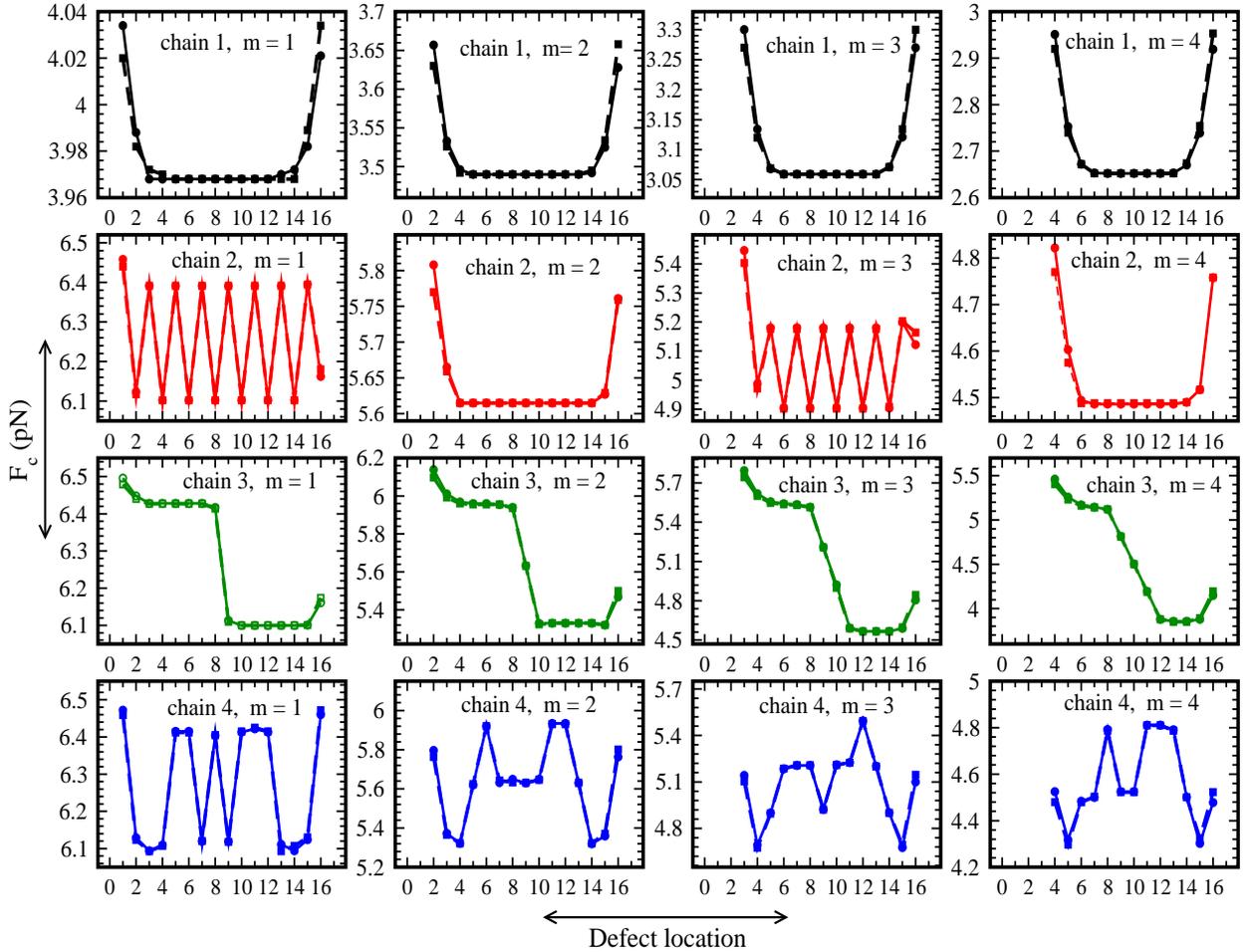}
\caption{\label{fig05} (Color online)
The value of critical force, $F_c$, for different chains having 
different number of defects as well as their location. We show both 
the cases when force is applied on $3'$-end (solid lines with circle) 
and on the $5'$-end (dashed lines with square). Only for the case when 
the defect site is in the middle of the chain is there no change in 
$F_c$ for these two case. The critical forces for pure 
\textit{ chains 1, 2, 3 \& 4} is 4.54, 6.96, 6.98, and 6.97 pN 
respectively.}
\end{center}
\end{figure*}
Let us consider {\it chain 1} with one defect.
As shown in the Fig. \ref{fig05} when the force is applied on the 
$3'$ -end and the defect pair is $1^{st}$ pair ($3'$ -end), the 
critical force reduces to 4.04 pN from 4.54 pN. This value further 
decreases to 3.99 pN when the defect pair is $2^{nd}$ pair. 
When the defect is located between 3-13 pairs, there is no change in 
the value of critical force, it is $\sim$3.97 pN. This means that
the base pair (defected) in this section of the chain have similar 
response to the applied force, irrespective of their location. 
The defect pair means a loop in the chain which will increase
the entropy of the chain. From the results, this is clear that
the loop contributes to the opening of the chain in addition
to the applied force and end entropy. However, 
as the defect location is somewhere between 14 to 16, 
contribution of bubble in the entropy of the chain is negligible.
Hence the critical force increases. In this case too, we observe a
necklace pattern. The pattern obtained here is not as symmetric as 
observed for thermal melting of the same chain with single defect. 
The reason for this difference lies in the nature of the chain opening in 
these two cases. We consider now the opening of the chain in another
condition. The force is applied on $3'$ -end and defect pair is the 
$5'$ -end. In this case, the critical force is 4.02 pN which is less
than for the previous case (where the force is on $3'$ -end and defect 
pair is also $3'$ -end) where $F_c$ is 4.04 pN. The difference is 
about 0.02 pN. The reason for this reduction is the difference 
in the end entropies for these two cases. In case when the defect 
end is $5'$- end, the entropy of this open end contributes to the opening. 
While for the first case, when the defect end is the $3'$ -end 
(the force is also on this end), the contribution from the $5'$ -end
will be less as it is an intact pair. Hence we need slightly higher 
force to open the chain for the first case. 
We obtained similar results for this chain with more defects ($m = 2,3,4$). For
all the investigations whenever $m>1$, all the defects are
consecutive defects. As the number of defects increases in the chain,
the difference in the $F_c$ for two different cases is greater.
In order to verify our arguments, we calculate the probabilities of 
opening of the base pairs for these two cases. 
The probability of opening of the $i^{th}$ pair, in a sequence is 
defined as \cite{ns2011}:
\begin{equation}
P_i = \frac{1}{Z_c}\int_{y_0}^{\infty} dy_i \exp\left[-\beta H(y_i, y_{i+1})\right] Z_j
\end{equation}
where 
\begin{equation}
 Z_j = \int_{-\infty}^{\infty} \prod_{j=1, j\neq i}^N dy_j 
 \exp\left[-\beta H(y_j, y_{j+1})\right]
\end{equation}
while $ Z_c $ is the configurational part of the partition function 
defined as in eq.\ref{eqn5}. For $y_0$, we have taken a value of 
2 ${\rm \AA}$.  To avoid the overflow of figures, we choose to display 
the surface plot for {\it chain 1} with 4 defects, see Fig. \ref{fig06}.
We observe that the difference in the critical force for these two
cases is $\sim 0.04$ pN.
\begin{figure*}[t]
\begin{center}
\includegraphics[height=3.0in,width=3.21in]{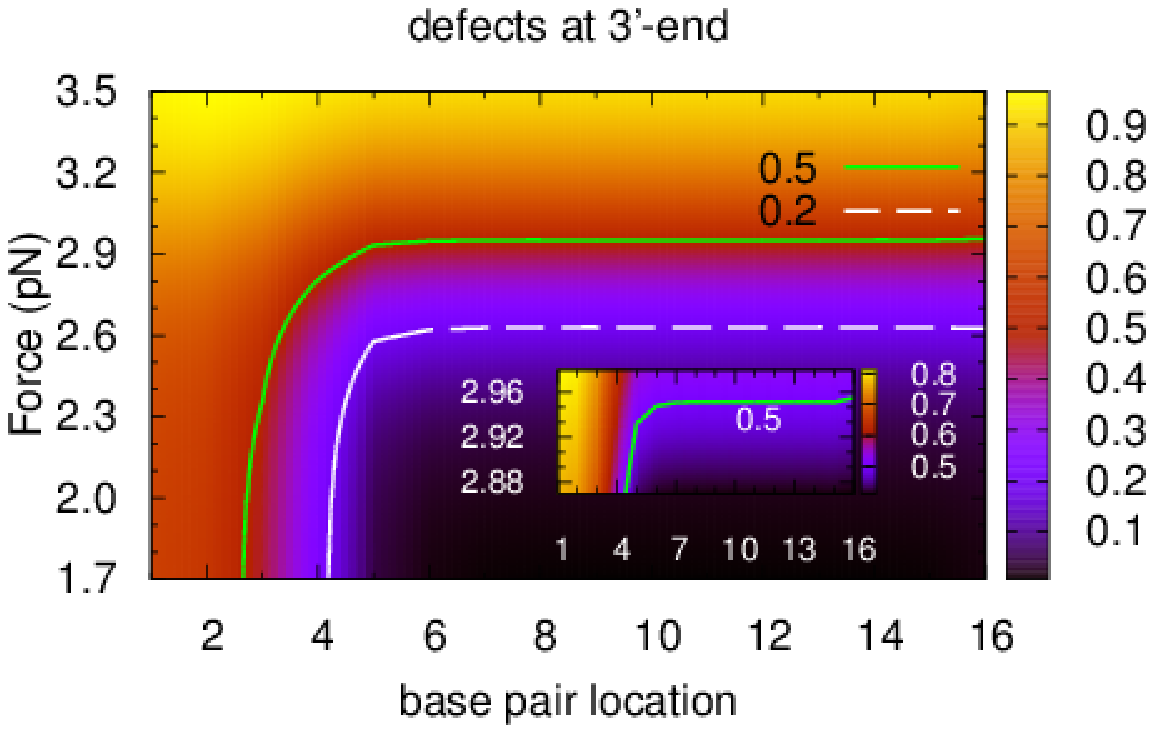}
\includegraphics[height=3.0in,width=3.21in]{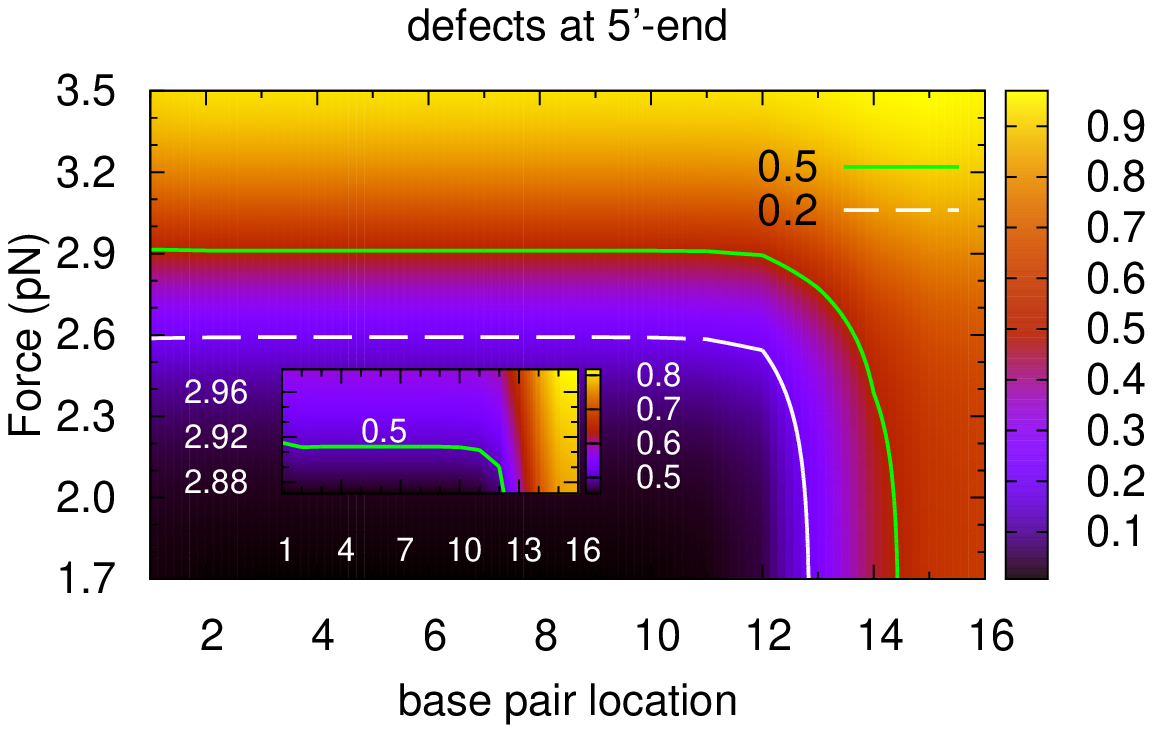}
\caption{\label{fig06} (Color online)
The density plots to show the difference in the opening of homogeneous 
DNA molecule ({\it chain 1 with four defects}) when force is applied on 
$3'$- end. (Left) When defect pairs are 1-4 ($3'$ end). 
(Right) When the four defect pairs are on the $5'$ end (13-16). 
The difference in the critical force $F_c$ for the two cases is 
observed here ({\it more clearly in the zoomed version}). 
In order to open 50\% of the base pairs, the $F_c$ for first case (left)
is 2.96 pN while for the second case (right) it is 2.92 pN. }
\end{center}
\end{figure*}

Let us consider, the {\it chain 2} (alternating $AT$ and $GC$ pairs).
In this case, we have four possible combinations of force and
defect locations. First, when force is applied on $3'$ -end
and the defect is also at the $3'$ -end. Second one is 
when force is applied on $3'$ -end and defect is at $5'$ -end. The other two cases
are the alternate combinations of these two. If we fix the location
of applied force at $3'$ -end and change the defect locations,
we find that for single defect the difference in $F_c$ is $\sim 0.3$ pN.
This is because of the difference in the entropy contribution from 
the two ends. In one case the end is $AT$ while in another case it is $GC$.
If we fix the defect location and change the applied force locations
from $3'$ -end to $5'$ -end, we find that the difference in 
$F_c$ is $\sim 0.02$ pN. The same argument which we gave for {\it chain 1}
with single defect is valid here too. Now consider this 
chain with two defects. When the force is applied on $3'$ -end and the
defect locations are $3'$ -end and $5'$ -end, the difference in $F_c$
is $\sim 0.04$ pN. This chain with two defects can be thought of as a 
homogeneous chain (of $AT + GC$ block) with single defect. However, 
the ends in this chain can be either $AT$ or $GC$ and hence we get a 
different pattern at the ends as compare to {\it chain 1}. 
Similar kind of feature is observed for the same chain with four defects. 
For the {\it chain 3}, 
the difference in the energy of $AT$ and $GC$ pair is clearly visible.
In this case, this is important on which end the force is applied. 
When the force is applied on $3'$ -end and the defect locations are 
$3'$ -end and $5'$ -end the difference in the $F_c$ is $\sim 0.32$ pN. 
The \textit{chain 4} is a chain with random distribution of $AT$ and $GC$
pairs. Since the distribution is random the energy landscape of $AT$ and
$GC$ pairs will play an important role in the opening of chain with different
locations of defect. The unzipping behavior of this chain displays some of
the features of all the three chains that we considered above. 
In case when the defect(s) are in the middle of the chain, 
the change in the value of critical force is negligible, {\it i.e.}, it 
does not matter from which end the chain is pulled.

\section{Conclusions}
\label{result}

In the present work, we have studied the role of defect(s) on the thermal
as well as mechanical denaturation of DNA molecule. It is known that
the defects delay the replication process which may further cause 
the cell death and hence may lead to initiation of cancer. 
Motivated by the experimental studies, we considered four 
different kind of DNA molecules. These molecules have different
numbers of \textit{AT} and \textit{GC} pairs and the distribution of
these pairs along the chain is also different. We have considered all the chains with 
$m$ number of defects, where $m$ varies from 1-4. Here we assumed that for
$m>1$ all the defects are in a block. For the equilibrium calculations,
we used PBD model and found the denaturation point in thermal as well
as in constant force ensembles. For the homogeneous chain, we found that 
there is a segment (4-12) of the chain where $T_m$ is unaffected by the 
location of the defect in the chain. In case of heterogeneous
chain, there is no plateau but it matters on a location whether there is
an \textit{AT} pair or a \textit{GC} pair. When we compared the opening in two
ensembles for homogeneous chain we observed that there is a striking difference.
While for the homogeneous chain we obtained a symmetric necklace kind of
plot in thermal ensemble, this was missing in force ensemble. This validates
the role of finite end entropy of the homogeneous chain in the denaturation
of the DNA molecule. For the thermal melting the ends have less
impact on the opening because of the fact that each base pair shares the 
same amount of thermal energy. There only the sequence of $AT/GC$ pairs 
matters. 

For the chain that is pulled from an end by some force, it is important 
for all kinds of chains (with defect) whether the force is applied on 
$3'$- end or $5'$- end. For unzipping in constant force ensemble we 
considered four possible cases. First two are when force is applied 
on $3'$- end and the defect locations are either on $3'$- or $5'$- ends.
Similarly other two combinations are when the force is applied on $5'$- end 
and defect locations are either $3'$- or $5'$- ends. In all these
cases, the nature of end pair is important. For the chain with 
alternate $AT$ and $GC$ sequence we observed that in addition to the 
ends the interface of defect and intact pair affect the opening of the 
chain. The interfaces for this chain with two defects are either of 
$AGA$ or $GAG$ kind. Hence there is a difference in the critical force
for the four cases. To show the importance of ends in the 
opening we calculated the probabilities of opening for the 
homogeneous chain with four defects. Here we considered two cases;
one when the force is applied on $3'$- end and defects are either at
$3'$- end or at $5'$- end. When the defect location is $3'$- end,
the end entropy is suppressed and hence we obtained a slightly higher critical
force for this case. The studies on \textit{chain 4}
are closer to the real chain as it has random sequence of
\textit{AT} and \textit{GC} pairs. The force profile (Fig. \ref{fig05})
shows the weak and strong sections of the chain. As a future of this work, 
one can study the opening of DNA molecule in both the ensemble as a 
function of time and the exact delay in the opening can be predicted. 
This is an attempt to understand the defect and their effect on the 
replication process. However, the real picture would be clearer  
from non-equilibrium studies.
How the cell decides which segment of DNA with a mismatch in the sequence can be
repaired or which would be destroyed will be an interesting area of
future studies. Is there any role of free energies of the sequence?
The time evolution of this kind of molecule may provide some
useful information.

\section*{Acknowledgement}
We are thankful to Y. Singh and S. Kumar, Department of Physics, Banaras Hindu
University, India, for useful discussions. We acknowledge the financial support 
provided by Department of Science and Technology, New Delhi [SB/S2/CMP-064/2013]
and University Grant Commission, New Delhi, India for BSR fellowships to AS.


\begin{thebibliography}{100}
\small{
\bibitem{watson}
J.D. Watson, F.H.C. Crick, Nature {\bf 171}, 737 (1953).
\bibitem{stryer}
L. Stryer, {\it Biochemistry}, Freeman, New York 1995.
\bibitem{pb}
M. Peyrard and A.R. Bishop, Phys. Rev. Lett. {\bf 62}, 2755 (1989); 
T. Dauxois, M. Peyrard and A.R. Bishop, Phys. Rev. E {\bf 47}, R44 (1993).
\bibitem{campa} 
A. Campa and A. Giansanti, Phys. Rev. E {\bf 58}, 3585 (1998).
\bibitem{chen}
Yong-li Zhang, Wei-Mou Zheng, Ji-Xing Liu, and Y. Z. Chen, Phys. Rev. E
{\bf 56}, 7100 (1997)
\bibitem{saul}
S. Ares, N.K. Voulgarakis, K.\O{}. Rasmussen, A.R. Bishop,
{\it Phys. Rev. Lett.} {\bf 94}, 035504 (2005);
G. Kalosakas and S. Ares, {\it J Chem. Phys.}, {\bf 130}, 235104 (2009).
\bibitem{voul}
N.K. Voulgarakis, A. Redondo, A. R. Bishop, and K. \O{}. Rasmussen, 
{\it Phys Rev Lett}, {\bf 96}, 248101 (2006).
\bibitem{erp}
T.S. van Erp, S. Cuesta-L\'{o}pez, J.G. Hagmann, M. Peyrad, Phys. Rev. Lett.
{\bf 95}, 218104 (2005);
T.S. van Erp, S. Cuesta-L\'{o}pez, M. Peyrad, Eur. Phys. J. E {\bf 20},
421 (2006).
\bibitem{weber}
G. Weber,  N. Haslam, J.W. Essex and J. Neylon, J. Phys. Condens. Matter {\bf 21}, 
034106 (2009).
\bibitem{pbd2009}
M. Peyrard,  S. Cuesta-L\'{o}pez and G. James, J. Biol. Phys. {\bf 35 }, 73 (2009).
\bibitem{alex}
B. Alexandrov, N.K. Voulgarakis, K.\O{}. Rasmussen, A. Usheva and A.R. Bishop, 
{\it J. Phys: Conden. Matter}, {\bf 21}, 034103 (2009).
\bibitem{zoli}
M. Zoli, J. Theor. Biol. {\bf 354}, 95 (2014).
\bibitem{ns01}
N. Singh and Y. Singh, Phys. Rev. E {\bf 64}, 042901 (2001).
\bibitem{ns03}
N. Singh and Y. Singh, Eur. Phys. J. E {\bf 17}, 7 (2005).
\bibitem{gauth}
M.G. Gauthier, J. Herrick, and J. Bechhoefer, Phys. Rev. Lett. {\bf 104}, 218104 (2010).
\bibitem{paivi}
P$\ddot{\rm a}$ivi Peltom$\ddot{\rm a}$ki, J. of Clinical Oncology, {\bf 21}, 1174 (2003).
\bibitem{hensey}
C. Hensey, and J. Gauthier, Mechanisms of Development {\bf 69}, 183 (1997).
\bibitem{vilen}
M.M. Vilenchik and A.G. Knudson, Proc. Natl. Acad. Sci. USA {\bf 100}, 12871 (2003).
\bibitem{branzei}
D. Branzei and M. Foiani, Curr. Opin. Cell Biol. {\bf 17}, 568 (2005).
\bibitem{kaufmann}
W. K. Kaufmann, Carcinogenesis {\bf 31}, 751 (2010).
\bibitem{mccullagh}
M. McCullagh, I. Franco, M.A. Ratner, G.C. Schatz, J. Phys. Chem. Lett. {\bf 3}, 689 (2013).
\bibitem{kafri}
Y. Kafri, D. Mukamel and L. Peliti, {\it Phys. Rev. Lett.}, {\bf 85}, 4988 (2000); {\it ibid}
{\it Eur Phys J. B}, {\bf 27}, 135 (2002).
\bibitem{amb}
T. Ambj\"{o}rnsson and R. Metzler, Phys. Rev. E {\bf 72}, 030901(R) (2005).
\bibitem{joyuex}
S. Buyukdagli, M. Joyeux, Phys. Rev. E {\bf 77}, 031903 (2008).
\bibitem{dudu}
C.I. Duduiala, J.A.D. Wattis, I.L. Dryden, C.A. Laughton, 
Phys. Rev. E {\bf 80}, 061906 (2009).
\bibitem{kumar} 
S. Kumar and M. S. Li, Phys. Rep {\bf 486}, 1 (2010).
\bibitem{frank}
M. D. Frank-Kamenetskii, Shikha Prakash, Physics of Life Reviews {\bf 11},
153 (2014).
\bibitem{ffalo}
A.E. Bergues-Pupo, J.M. Bergues, F. Falo, Physica A, {\bf 396}, 99 (2014);
A.E. Bergues-Pupo {\it et al}, EPL {\bf 105}, 68005 (2014).
\bibitem{macedo}
D.X. Macedo, I. Guedes, E.L. Albuquerque, Physica A {\bf 404}, 234 (2014).
\bibitem{nik}
Nikos Theodorakopoulos, Phys. Rev. E {\bf 82}, 021905(2010).
\bibitem{as_phy02}
A. Singh and N. Singh, Physica A: Statistical Mechanics and its Applications,
{\bf 419}, 328 (2015) and \textit{references therein}
\bibitem{pbd95}
T. Dauxois, M. Peyrard, Phys. Rev. E {\bf 51}, 4027 (1995)
\bibitem{somen}
Garima Mishra, Poulomi Sadhukhan, Somendra M. Bhattacharjee, and Sanjay Kumar, 
Phy. Rev. E {\bf 87}, 022718 (2013);
Poulomi Sadhukhan, Somendra M. Bhattacharjee, arXiv:1401.5451.
\bibitem{hatch}
K. Hatch, C. Danilowicz, V. Coljee and M. Prentis, Nucleic Acids Res. {\bf 36}, 294 (2008).  
\bibitem{huguet}
Josep M. Huguet {\it et al}, Proc. Natl. Acad. Sci., USA {\bf 107}, 15431 (2010).
\bibitem{ritort}
F. Ritort, J. Phys. Condens. Mat. {\bf 18}, R531 (2006).
\bibitem{ns2011}
S. Srivastava and N. Singh, {\it J. Chem. Phys.}, {\bf 134}, 115102 (2011).
}
\end{thebibliography}
\end{document}